# Theoretical conditions of pipe flow transition: new approach to the old problem

Andrei Nechayev

*Moscow State University, Geographic Department*

*For Hagen-Poiseuille flow the mechanism of laminar-to-turbulent transition originated from the deceleration of disturbed streams on the pipe wall is analyzed. An expression for the critical Reynolds number and the disturbance in the form of ring obstacle is derived. The good qualitative agreement between the theoretical results and published experimental data is demonstrated*

Turbulence is a special state of the flowing fluid, which macroscopic particles moving parallel to the flow axis (i.e. laminar) begin to experience complex, sometimes chaotic, transverse and rotational motions. Literary description of turbulence and its vortex nature was given by Leonardo da Vinci. At the end of the nineteenth century Osborne Reynolds laid the foundation for turbulence research receiving critical condition for laminar-to-turbulent transition for the flow in the pipe, a condition which later became a classic. Since then, a lot of effort was made to build a simple and clear physical and mathematical model of turbulence and of the laminar-to-turbulent transition, but failed to do it.

It should seem to recognize that a purely mathematical approach to the description of turbulence does not prove its value. Attempt to analyze the laminar-to-turbulent transition on the basis of methods of stability theory was not successful, because with its help the critical Reynolds number in its simplest form for the standard flow in a pipe was not derived. Also the dependence of this Number from the heterogeneity amplitude introduced into the flow known from experiments was not demonstrated even qualitatively. Hagen-Poiseuille flow, as noted in [1], represents a serious challenge to the theory of hydrodynamic stability, as it is theoretically stable with respect to small fluctuations at any Reynolds number. However, even at Re > 4000 in the presence of small perturbation the stream becomes fully turbulent.

In this paper, we will continue to analyze the physical nature of the turbulence. We assume, as in [2], that a necessary (but not sufficient) condition for the transition to turbulence is the appearance in the laminar flow of individual streams directed against the current. It is known that the presence of counter-flow stimulates and accompanies the formation of vortex structures. Obviously, to generate the flow in the back direction the existing of reverse pressure gradient is necessary. To justify the emergence of such a gradient let's return back to the basics of classical hydrodynamics: to Bernoulli's equation. This equation imply, figuratively speaking, energetic "swing". The kinetic energy of the particles in a stationary fluid stream and its pressure are associated so that in the acceleration zones fluid pressure decreases and in the zones of deceleration pressure increases in accordance with the relation $\partial p/\partial s = -\rho v \partial v/\partial s$ where $v$ is a particle velocity in the flow direction $s$. This local increase in pressure can and should be observed near solid surfaces, walls of pipes where streams deviated from the axis of the flow are decelerated. In the zones of deceleration the kinetic energy of the particles transforms into pressure potential which can restore the same velocity of the particles but in a new direction.

Thus, the deceleration of disturbed streams of laminar flow on the walls of pipe or channel can cause backflow and formation of the corresponding vortex structures.

Reverse flow may occur if a local increase in pressure due to the deceleration of the stream exceeds the corresponding pressure drop in viscous flow. The maximum pressure increase in the decelerating zone of the stream is obviously $\rho v^2 / 2$ where $v$ is the disturbed stream linear velocity which is reduced to zero. The magnitude of this velocity will be determined by the nature of the disturbance introduced in the flow. In contrast to our previous work [2], where the disturbed jet collides with the nearest wall, we consider a perturbation similar to that used in the experimental work [3]. It was a circular obstacle or iris diaphragm placed in the fully developed Poiseuille flow. Figure 1 reproduces this situation, where the red color shows the expected trajectory of perturbed streams.

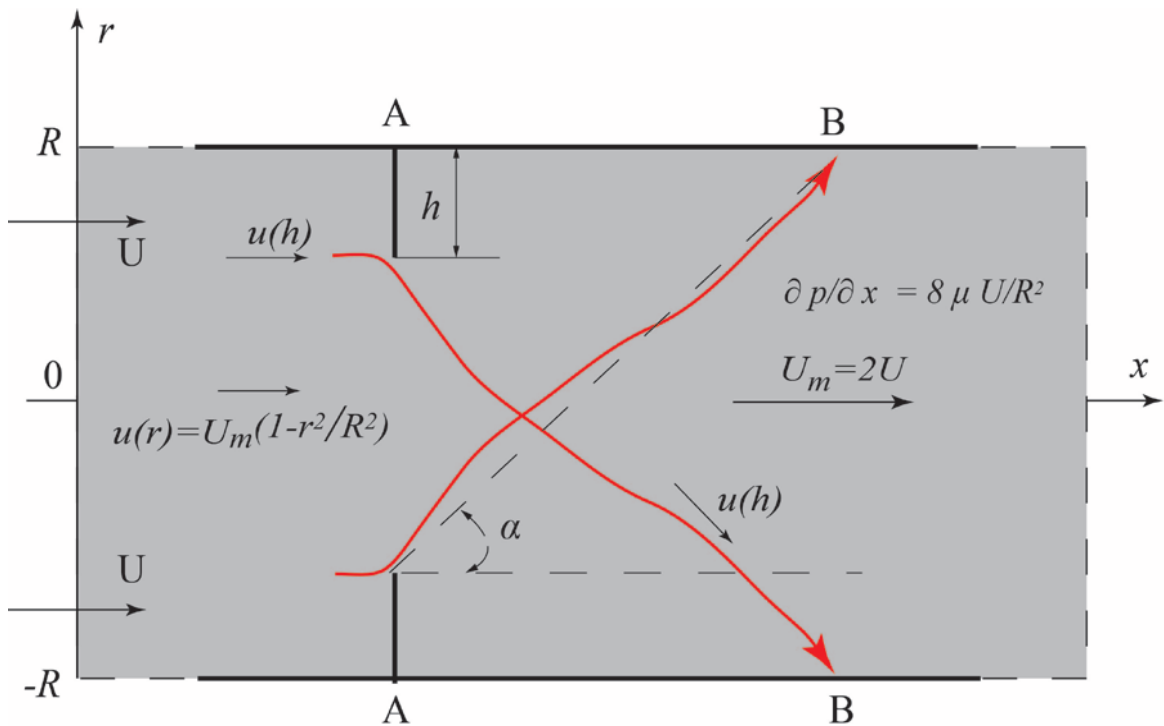

Figure 1. *Flow of Hagen-Poiseuille in a pipe with an obstacle in the form of an annular aperture. Red color indicates the disturbed streams.*

As the amplitude of the heterogeneity in [3] the parameter $h$ was used – it is the height of the ring or diaphragm protrusion above the surface of the wall (Fig.1). Since the annular obstacle is located in the zone of fully developed Poiseuille flow, the maximum velocity of the particles in the disturbed stream should not exceed *u (h)*. Deviation of the stream will be directed to the opposite side of the pipe (Fig. 1). Point B of collision with a solid surface is located at a distance of AB downstream. This distance is equal to *(2R-h) / tgα*, where α is a conditional deflection angle of the stream. We can assume that some part of the stream will be decelerated completely at point B on the roughness of the wall, and pressure increasing at this point will be $\rho u^2(h)/2$. Thus, the occurrence of backflow must be expected under the following conditions:

$$\rho u^2(h)/2 > \frac{\partial p}{\partial x} \frac{2R-h}{tg\alpha} \qquad (1)$$

Using the standard formula for Poiseuille flow $\{ u(r) = 2U(1-r^2/R^2); \partial p/\partial x = 8\mu U/R^2 \}$ and introducing the dimensionless parameter $k = h/R$, the condition (1) can be rewritten as:

$$\rho U^2 k^2 (2-k) > \frac{8\mu U}{D tg\alpha} \qquad (2)$$

where $U$ is the mean flow velocity, $D$ is the diameter of the pipe ($D = 2R$), $\mu$ - viscosity, $\rho$ - density. In its final form:

$$\mathrm{Re} \equiv \frac{\rho D U}{\mu} > \frac{8}{k^2(2-k)tg\alpha} \equiv \mathrm{Re}_{cr} \qquad (3)$$

Note that for small amplitudes of heterogeneity ($h << R$, $k << 1$) we obtain the same dependence $\mathrm{Re}_{cr} \sim h^{-2}$ which has been observed experimentally [3] for large value of the critical Reynolds number.

Thus, the transition to turbulence in our interpretation is due to the appearance of reverse streams in the perturbed laminar flow, and the critical Reynolds number is smaller the stronger stream deviation (more tgα), the greater its speed (the closer the disturbed stream to the axis of the flow). The presence of a counter-flow in the disturbed streams, leading to turbulence, is indirectly confirmed by visual observation [6] in the pipe flow after the injection of the dyed jet.

As follows from (3), the minimum value of the Reynolds number required for the emergence of reverse flow may be less than 8 when the central stream is subjected to the perturbation. But the single vortex may be insufficient for the turbulence development, unless the vortex begins to disturb the central part of the flow, deflecting streams with velocities larger than $u(h)$ for which the critical condition (3) is automatically satisfied. In this case the process of vortex formation can occur as avalanche which provides a transition to turbulent flow. If condition (3) is not satisfied there is no counter-flow streams and the deviation decays downstream.

Our approach allows at least a qualitative explanation of the physic origin of the two major turbulent structures, puffs and slugs [4]. First occur at relatively low values of Reynolds number and large perturbations which, in accordance with our model, act on the central streams of laminar flow with high velocities. The transformation of these streams in the vortex structures under decelerating on the pipe wall should disturb the flow but away from its axis, where the fluid velocity is less and the critical condition (3) is not satisfied. So the primary flow will not be "picked up" and dissipates downstream (as it happens with the "puffs" at Re <2000). If the perturbation undergone stream near the wall (low-speed and small amplitude of heterogeneity) and (3) is satisfied, the formation of the primary vortex and its drift to the center will disturb streams with high velocities for which the condition (3) is satisfied, so it will launch "chain reaction" of the vortices formation and result in turbulent "slug". Ideas expressed here may shed light on the origin of the minimum critical Reynolds number.

Using the above approach, stationary, self-sustaining existence of turbulent structures can be explained as follows. Initial heterogeneity disturbs some stream, which is decelerated at the wall, turns completely or partially in the opposite direction, since (3) is satisfied, and forms a vortex structure, which triggers the formation of other vortex structures. These structures should have a drift velocity significantly less than the maximum flow rate, since the principle of their formation (collision of counter-flows and twisting) implies the conservation of the initial total momentum close to zero (counter-flow streams with the same module of speed), as well as the conservation of the kinetic energy of the linear movement in the form of rotational energy. Thus, the generated vortices will be a factor for the deceleration of the central streams, performing the function of primary heterogeneity. These central streams will be deflected to the walls, decelerating and triggering the formation of new vortices. A similar structure of the "puff" was described in [5].